\documentclass[shortbibliography,twocolumn,prl,aps,superscriptaddress,amsmath,amssymb,floatfix]{revtex4-2}
\usepackage{amsfonts}
\usepackage{mathrsfs}
\usepackage{amsmath}
\usepackage{xcolor}
\usepackage{graphicx}
\usepackage{bm}
\usepackage{amssymb}
\usepackage{xspace}
\usepackage{epstopdf}
\usepackage{dcolumn}
\usepackage{longtable}
\usepackage{multirow}
\usepackage{float}
\usepackage{comment}
\usepackage{varwidth}
\makeatletter

\usepackage[colorlinks=true, letterpaper=true, pdfstartview=FitV, linkcolor=blue, citecolor=blue, urlcolor=blue]{hyperref}
\makeatother

\begin{document}

\title{Nonlinear N\'eel Spin-Orbit Torque in Centrosymmetric Antiferromagnets}

\author{Jin Cao}
\thanks{These authors contributed equally to this work.}
\affiliation{Institute of Applied Physics and Materials Engineering, Faculty of Science and Technology, University of Macau, Macau, China}

\author{Weikang Wu}
\thanks{These authors contributed equally to this work.}
\affiliation{Key Laboratory for Liquid-Solid Structural Evolution and Processing of Materials, Ministry of Education, Shandong University, Jinan 250061, China}

\author{Huiying Liu}
\email{liuhuiying@buaa.edu.cn}
\affiliation{School of Physics, Beihang University, Beijing 100191, China}

\author{Shen Lai}
\affiliation{Institute of Applied Physics and Materials Engineering, Faculty of Science and Technology, University of Macau, Macau, China}

\author{Cong Xiao}
\email{congxiao@fudan.edu.cn}
\affiliation{Interdisciplinary Center for Theoretical Physics and Information Sciences (ICTPIS), Fudan University, Shanghai 200433, China}

\author{X. C. Xie}
\affiliation{Interdisciplinary Center for Theoretical Physics and Information Sciences (ICTPIS), Fudan University, Shanghai 200433, China}
\affiliation{International Center for Quantum Materials, School of Physics, Peking University, Beijing 100871, China}

\author{Shengyuan A. Yang}
\affiliation{Research Laboratory for Quantum Materials, Department of Applied Physics, The Hong Kong Polytechnic University, Hong Kong, China}

\begin{abstract}
Electric control of N\'eel vector is a central task of antiferromagnetic (AFM) spintronics. The major scheme so far relies on the linear N\'eel torque, which however is restricted to AFMs with broken inversion symmetry. Here, we propose a nonlinear N\'eel spin-orbit torque, uniquely enabling electric control in the vast class of centrosymmetric AFMs, where the existing scheme fails. Importantly, its intrinsic component, rooted in sublattice-resolved band quantum geometry, offers two additional advantages: It operates also in
$\mathcal{PT}$-symmetric AFM insulators, where linear torque is forbidden; and it has anti-damping character, making it more efficient in driving magnetic dynamics. Combined with first-principles calculations, we predict large effect in MnRh and MnBi$_{2}$Te$_{4}$, which can be readily detected in experiment.
Our work unveils a new fundamental effect, offers a new strategy of electric control in AFM systems beyond the
existing paradigm, and opens the door to the field of nonlinear AFM spintronics.

\end{abstract}

\maketitle

\begin{figure}
\begin{centering}
\includegraphics[width=8.6cm]{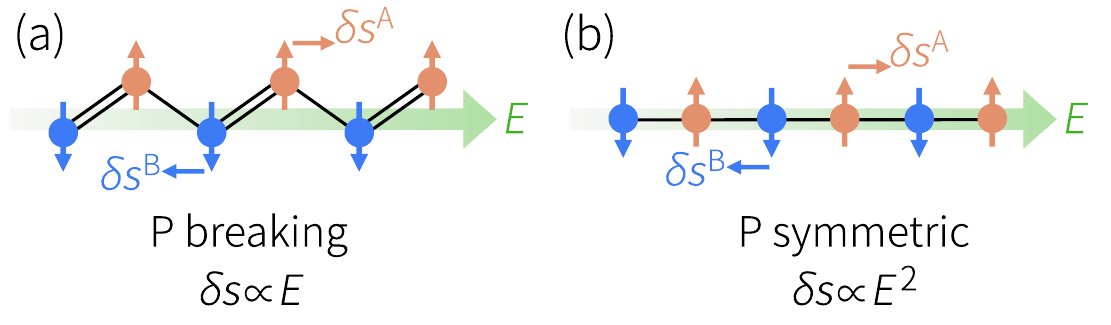}
\par\end{centering}
\caption{\protect\label{Fig_1}Schematics of (a) linear N\'eel torque due to 
linearly field-induced staggered spin polarization in AFMs with broken inversion symmetry; and (b) 
nonlinear N\'eel torque that can operate in centrosymmetric AFMs, where the linear N\'eel torque is forbidden.}
\end{figure}

The potential of antiferromagnets (AFM) in spintronics applications has received tremendous interest, due to their unique advantages, such as robustness against magnetic field perturbations, absence of stray fields, and ultrafast spin dynamics~\citep{Jungwirth2016Antiferromagnetic,Baltz2018Antiferromagnetic,Han2023Coherent}. To make scalable devices, a central task of AFM spintronics is to realize the electric control of N\'eel order~\citep{Manchon2019Current}. Currently, the main approach  is via the so-called N\'eel torque~\citep{Zelezny2014Relativistic,Zelezny2017Spin,Zelezny2018Spin}, exerted by electrically induced local spin polarization that is \emph{staggered} on opposite AFM spin sublattices, as illustrated in Fig.~\ref{Fig_1}(a). The  N\'eel torque and its effect on N\'eel vector dynamics have been successfully demonstrated in several AFMs, including CuMnAs~\citep{Wadley2016Electrical,Grzybowski2017Imaging,Godinho2018Electrically} and Mn$_{2}$Au~\citep{Bodnar2018Writing,Jourdan2023Mn2Au}, with a greatly improved operation speed up to GHz~\citep{Olejnik2017Antiferromagnetic} or even THz scale~\citep{Olejnik2018Terahertz}.


Despite significant progress, the current scheme of N\'eel torque relies on linear effects and is restricted to AFMs with broken inversion symmetry $\mathcal{P}$ ~\citep{Zelezny2018Spin}. Indeed, in a centrosymmetric AFM, the $\mathcal{P}$-invariance of magnetic moment necessitates that each magnetic sublattice also preserves $\mathcal{P}$. Consequently, the linearly induced spin polarization $\delta s\propto E$, being odd under $\mathcal{P}$, must have a zero net value on each sublattice, resulting in a vanishing N\'eel torque. Given that centrosymmetric AFMs constitute a very large class of AFM materials with rich functionalities, how to achieve N\'eel torque in these systems has become a critical challenge for AFM spintronics~\cite{Jungwirth2016Antiferromagnetic,Baltz2018Antiferromagnetic,Han2023Coherent,Zelezny2018Spin}.

%
%


In this work, we address this challenge by proposing a \emph{nonlinear} N\'eel torque, driven by sublattice-staggered spin polarization nonlinearly generated by electric field [Fig.~\ref{Fig_1}(b)]. This is a general effect working for both noncentrosymmetric and centrosymmetric AFMs. From symmetry analysis, we show that for centrosymmetric AFMs, it is the leading order mechanism for N\'eel torque; and for certain noncentrosymmetric AFMs, e.g., $\mathcal{PT}$-symmetric AFM insulators, it also gives the dominant contribution and is of entirely intrinsic nature. Moreover, the intrinsic part of 
nonlinear N\'eel torque has anti-damping character, making it highly efficient in driving magnetic reorientations.
We develop a microscopic theory for the intrinsic nonlinear
N\'eel torque and reveal its origin in sublattice-resolved band geometry.  
Combining with first-principles calculations,
we evaluate the proposed effect in MnRh, 3D bulk MnBi$_2$Te$_4$, and 2D even-layered MnBi$_2$Te$_4$, finding significant results that
can be experimentally detected. Our work
not only reveals a fundamental phenomenon in nonlinear spintronics, but also opens a new route to electrical manipulation of AFM order in a wide range of previously unaccessible systems, paving the way towards nonlinear AFM spintronics


\textcolor{blue}{\textit{Symmetry consideration.}} Consider fully compensated collinear AFM systems with two magnetic sublattices, labeled by $i=A,B$.
The nonlinear spin polarization generated by applied $E$ field on sublattice $i$ can be expressed as
\begin{equation}\label{ds}
  \delta s_{a}^{i}=\alpha_{abc}^{i}E_{b}E_{c}
\end{equation}
where $\alpha_{abc}^{i}$ is the sublattice-resolved nonlinear response tensor, and the summation over repeated Cartesian indices (the subscripts here) is implied henceforth. One immediately observes that distinct from linear N\'eel torque, due to the additional $E$ factor, $\delta s^i$ here is even under $\mathcal{P}$, making the nonlinear response compatible with centrosymmetric AFMs.

The specific form of $\alpha$ tensor are constrained by symmetry group of AFMs.  To analyze sublattice-resolved responses, we make the following decomposition of a given magnetic point group $\mathcal{G}$:
\begin{equation}
    \mathcal{G}=\mathcal{G}_s+a\mathcal{G}_s,
\end{equation}
%
%
%
where $\mathcal{G}_s$ is the subgroup of operations that preserve each sublattice, and $a\mathcal{G}_s$ are the operations that switch the two sublattices, with $a$ being any representative of this set. Moreover, in studying magnetic systems, it is important to distinguish the character of a response under time reversal operation $\mathcal{T}$, i.e., whether the response changes sign under the reversal of all magnetic moments. Any response tensor can be decomposed into $\mathcal{T}$-odd and $\mathcal{T}$-even parts. For a $\mathcal{T}$-odd response $\alpha_{abc}^{i}$, it is constrained by
%
%
\begin{equation}
\begin{cases}
\alpha_{a^{\prime}b^{\prime}c^{\prime}}^{i}=\eta\left|\mathcal{O}\right|\mathcal{O}_{a^{\prime}a}\mathcal{O}_{b^{\prime}b}\mathcal{O}_{c^{\prime}c}\alpha_{abc}^{i}, & \mathrm{if}\,\mathcal{O}\in \mathcal{G}_s\\
\alpha_{a^{\prime}b^{\prime}c^{\prime}}^{A}=\eta\left|\mathcal{O}\right|\mathcal{O}_{a^{\prime}a}\mathcal{O}_{b^{\prime}b}\mathcal{O}_{c^{\prime}c}\alpha_{abc}^{B}, & \mathrm{if}\,\mathcal{O}\in a\mathcal{G}_s
\end{cases}, \label{symmetry}
\end{equation}
where $\left|\mathcal{O}\right|=\pm 1$ is the determinant of $\mathcal{O}$, $\eta=-1$ for primed operations, i.e., operations involving $\mathcal{T}$ operator, and $\eta=1$ for nonprimed operations. The constraint (\ref{symmetry}) also applies to $\mathcal{T}$-even $\alpha_{abc}^{i}$ response, for which $\eta=1$ for both primed and nonprimed operations.

To support a N\'eel torque, we must have: (i) $\mathcal{G}_s$ allows components of $\delta\boldsymbol{s}^{i}$ perpendicular to the sublattice magnetization $\bm M^i$; and (ii) $a\mathcal{G}_s$ allows staggered spin polarization on the two sublattices, i.e., $\delta\boldsymbol{s}^{A}=-\delta\boldsymbol{s}^{B}$.
Based on Eq.~(\ref{symmetry}), we examine the constraints from all magnetic symmetry elements on $\alpha_{abc}^{i}$, and the results are presented in Supplemental Material~\cite{sm}.

Here, we highlight some important consequences from this analysis. First, many previous studies were focusing on AFMs with $\mathcal{PT}$ or $\mathcal{T}t_{1/2}$ symmetry ($t_{1/2}$ is a half lattice translation). We note that under  $\mathcal{PT}$, the linear N\'eel torque must be a $\mathcal{T}$-even response, however, the nonlinear torque is a
$\mathcal{T}$-odd response. This is a crucial difference, as it determines the different character of the torque: The $\mathcal{T}$-odd response may correspond to an antidamping-like torque, whereas the $\mathcal{T}$-even response tends to give a field-like torque, and the two act on magnetic dynamics in qualitatively different ways \cite{Manchon2019Current}. Meanwhile, we also note
that for AFM with $\mathcal{T}t_{1/2}$ symmetry, linear and nonlinear N\'eel torques are both of $\mathcal{T}$-odd character. These results are presented in Table~\ref{tab1}.

Second, there exist a number of AFMs, e.g., $L1_{0}$-type manganese alloys, NiO, and 3D MnBi$_2$Te$_4$, which possess both $\mathcal{PT}$ and $\mathcal{T}t_{1/2}$ symmetries (This implies $\mathcal{P}$ is also preserved). According to Table~\ref{tab1}, in such systems,  linear N\'eel torque as well as $\mathcal{T}$-even nonlinear N\'eel torque are all prohibited, and the dominating torque is from the $\mathcal{T}$-odd nonlinear mechanism.


Third, the nonlinear N\'eel torque may also play a leading role in certain noncentrosymmetric AFMs. For example, in $\mathcal{PT}$-symmetric AFM \emph{insulators} (regardless of whether $\mathcal{P}$ is respected or not), linear N\'eel torque is always suppressed, because it is $\mathcal{T}$-even and of extrinsic origin (i.e., arising from scattering effects), which requires the presence of Fermi surface~\cite{Zelezny2014Relativistic,Zelezny2017Spin}. Consequently, the nonlinear response also dominates such cases, and it is of entirely intrinsic nature, determined by the intrinsic quantum geometry of valence electrons (see below), rendering a new way to manipulate $\mathcal{PT}$-symmetric AFM insulators.


\begin{table}
\centering
\caption{\label{tab1} Comparison of linear and nonlinear N\'eel torques in several commonly encountered AFM systems. Here, BCP indicates the nonlinear torque in insulating AFMs is entirely from the intrinsic BCP mechanism.}
\begin{ruledtabular}
\renewcommand{\arraystretch}{1.2}
\begin{centering}
\begin{tabular}{ccccc}
 & \multicolumn{2}{c}{Linear} & \multicolumn{2}{c}{Nonlinear}\tabularnewline
 & $\mathcal{T}$-even & $\mathcal{T}$-odd & $\mathcal{T}$-even & $\mathcal{T}$-odd\tabularnewline
\hline
$\mathcal{P}$ & $\times$ & $\times$ & $\checkmark$ & $\checkmark$\tabularnewline
$\mathcal{PT}$ & $\checkmark$ & $\times$ & $\times$ & $\checkmark$\tabularnewline
$\mathcal{T}t_{1/2}$ & $\times$ & $\checkmark$ & $\times$ & $\checkmark$\tabularnewline
Insulators & $\times$ & $\checkmark$ & $\times$ & $\checkmark$ (BCP)\tabularnewline
\end{tabular}
\par\end{centering}
\end{ruledtabular}
\end{table}

\textcolor{blue}{\textit{Intrinsic mechanism.}} We shall focus on intrinsic nonlinear N\'eel torque, based on the following considerations. First, this intrinsic nonlinear torque is $\mathcal{T}$-odd, so it is relevant to all the important cases in Table~\ref{tab1}. Particularly, for most collinear AFM insulators (with $\mathcal{P}$ or $\mathcal{PT}$ symmetry), it is the \emph{only} contribution present.
Second, the intrinsic nonlinear N\'eel torque is determined solely by the band structure, manifesting the inherent property of each material. Hence, it can be quantitatively evaluated and provides a benchmark for comparison between theory and experiment.


Based on the extended semiclassical theory~\cite{Gao2014Field,Gao2015Geometrical,Dong2020Berry,Xiao2021OM}, the intrinsic nonlinear electrically-induced spin polarization
has been derived in ferromagnetic systems~\cite{Xiao2022Intrinsic}. To deal with N\'eel torques in AFMs, we need to distinguish the two magnetic sublattices and formulate the sublattice-resolved response tensor $\alpha_{abc}^{i}$ (details in \cite{sm}). The following formula is obtained (we set $e=\hbar=1$):
%
\begin{eqnarray}
\alpha_{abc}^{i} & = & -\frac{1}{2}\int [d\bm k]f_{0}^{\prime}\left(s_{a}^{i}G_{bc}+v_{b}\mathfrak{G}_{ac}^{i}+v_{c}\mathfrak{G}_{ab}^{i}\right)\nonumber \\
 &  & -\frac{1}{2}\int [d\bm k]f_{0}\partial_{h_{a}^{i}}G_{bc}\Big|_{h^i=0},\label{alpha}
\end{eqnarray}
where we have explicitly symmetrized the two subscripts $b$ and $c$, $[d\bm k]$ is a shorthand for $\sum_n d k^D /(2\pi)^D$ with $D$ the spatial dimension of the system, the band index label $n$ and the $k$ dependence of quantities in the integrand  are not written out explicitly for simple notations. $f_{0}$ is the Fermi-Dirac distribution, ${s}_{a}^{i}=\langle u_{n\bm k}|\hat{s}_{a}^{i}|u_{n\bm k}\rangle$ is the sublattice-resolved spin polarization for Bloch state $|u_{n\bm k}\rangle$,
with $\hat{s}_{a}^{i}=\frac{1}{2}(\hat{s}_{a}\hat{P}_{i}+\hat{P}_{i}\hat{s}_{a})$ and $\hat{P}_{i}$ the projection operator for sublattice $i$, and $v_a$ is the group velocity. $h_{a}^{i}$ is an auxiliary field coupled to sublattice spins through a term $-s_{a}^{i}h_{a}^{i}$~\cite{Dong2020Berry} added to the Hamiltonian and it is taken to be zero at the end of calculation (the explicit expression of $\partial_{h^{i}}G$ is presented in \cite{sm}). There are two important band geometric quantitites appearing in Eq.~(\ref{alpha}).
$G_{ab}$, known as the momentum-space Berry-connection polarizability (BCP)~\cite{Liu2021NAHE}, can be expressed in terms of interband velocity matrix element $\left(v_{a}\right)^{nn^{\prime}}$ and band energy $\varepsilon_{n}$:
\begin{equation}
  G_{ab}=2\mathrm{Re}\sum_{n^{\prime}\neq n}\frac{\left(v_{a}\right)^{nn^{\prime}}\left(v_{b}\right)^{n^{\prime}n}}{\left(\varepsilon_{n}-\varepsilon_{n^{\prime}}\right)^{3}}.
\end{equation}
Meanwhile, $\mathfrak{G}_{ab}^{i}$ is the sublattice-resolved $h$-space BCP:
\begin{eqnarray}
\mathfrak{G}_{ab}^{i} & = & -2\mathrm{Re}\sum_{n^{\prime}\neq n}\frac{\left(s_{a}^{i}\right)^{nn^{\prime}}\left(v_{b}\right)^{n^{\prime}n}}{\left(\varepsilon_{n}-\varepsilon_{n^{\prime}}\right)^{3}}.
\end{eqnarray}

We have a few remarks on this result. First, one observes that the intrinsic response contains both a Fermi surface contribution (the first line of Eq.~(\ref{alpha})) and a Fermi sea contribution (the second line).
The Fermi surface term is present only for metallic AFMs, while the Fermi sea term can operate also in insulating AFMs.
Second, one can verify that the result of (\ref{alpha}) is even under $\mathcal{P}$ (and odd under $\mathcal{T}$), confirming its compatibility with centrosymmetric AFMs. It should be noted that although Eq.~(\ref{alpha}) gives the
nonlinear electric spin generation on the two sublattices, to obtain the N\'eel torque, one needs to extract its
staggered component, i.e., the antisymmetric part with $\alpha^{A}=-\alpha^{B}$. Notably, for AFMs with suitable symmetries, e.g.,
$\mathcal{PT}$ or $\mathcal{T}t_{1/2}$ symmetry, $\alpha$ given by Eq.~(\ref{alpha}) is constrained to be staggered on the sublattices hence directly gives the N\'eel torque.
Third, the BCPs are typically pronounced near band (anti)crossing regions in a band structure. This offers a useful guidance to enhance the nonlinear N\'eel torque.


As mentioned, our proposed nonlinear effect could be the dominant source of N\'eel torque for a large class of AFM materials. Below, we apply our theory to three representative examples, including the centrosymmetric AFMs MnRh and bulk MnBi$_2$Te$_4$, and $\mathcal{PT}$-symmetric even-layer MnBi$_2$Te$_4$.

\begin{figure}
\begin{centering}
\includegraphics[width=8.6cm]{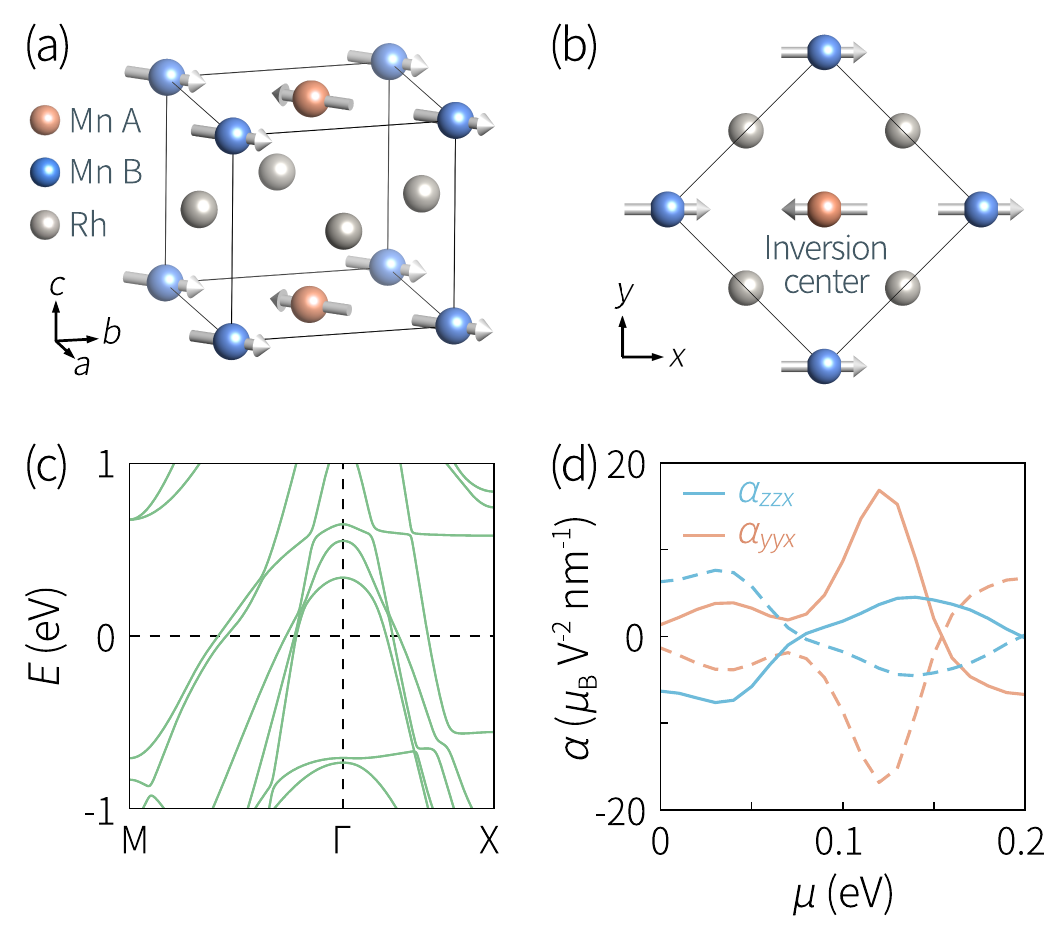}
\par\end{centering}
\caption{\protect\label{Fig_2}(a) Structure of $L1_{0}$-type AFM MnRh. (b) Top view of (a). (c) Calculated band structure of MnRh. (d) Calculated $\alpha_{yyx}^{i}$ and $\alpha_{zzx}^{i}$ near the Fermi level. Solid (dashed) lines correspond to $i=A$ ($B$).  $T=100$~K is taken in the calculation of (d).}
\end{figure}

\textcolor{blue}{\textit{Application to MnRh.}} Manganese alloys form a large family of magnetic materials~\cite{Fukamichi2006chapter}. Most of them are AFM with high N\'eel temperature and possess inversion symmetry. Here, we consider  $L1_{0}$-type MnRh, which has a tetragonal structure with $D_{4h}$ point group (Fig.~\ref{Fig_2}(a)).
In its AFM ground state, the N\'eel vector is in-plane and  along the {[}110{]} direction~\cite{Pal1968Magnetic,Khmelevskyi2011Element,Park2019Strain}.
Our first-principles calculations (details in \cite{sm}) find the magnetic moment at Mn site is  $\sim 3.1\,\mu_{\mathrm{B}}$, consistent with previous study~\cite{Park2019Strain}.

The magnetic point group of MnRh is $mmm1^{\prime}$, preserving both $\mathcal{P}$ and $\mathcal{PT}$ (hence also $\mathcal{T}t_{1/2}$) symmetries. According to Table~\ref{tab1}, the linear response is forbidden, and the N\'eel torque is dominated by
$\mathcal{T}$-odd nonlinear response which is enforced to be sublattice-staggered.
In the coordinate system of Fig.~\ref{Fig_2}(b), our symmetry analysis shows that
N\'eel torque is associated with two independent response coefficients $\alpha_{yyx}^{i}$ and $\alpha_{zzx}^{i}$.


Figure~\ref{Fig_2}(c) shows the calculated band structure of MnRh, which is an AFM metal. The intrinsic $\alpha_{yyx}^{i}$ and $\alpha_{zzx}^{i}$ determined by the band structure are computed and plotted versus chemical potential in Fig.~\ref{Fig_2}(d).
As mentioned, the N\'eel vector of this system tends to be in-plane: the in-plane magnetic anisotropy energy (MAE) $\sim$ a few $\mu$eV is much smaller than the out-of-plane MAE $\sim 0.1$ meV~\cite{Khmelevskyi2011Element,Park2019Strain}. Hence, it should be easier to rotate the N\'eel vector in $xy$-plane. Consider $\alpha_{zzx}^{i}$, which takes a value of $6\,\mu_{\mathrm{B}}\,\mathrm{V^{-2}\,nm^{-1}}$ at intrinsic Fermi level. Under a typical current density of $10^8~\mathrm{A/cm^{2}}$~\cite{Olejnik2018Terahertz,Klaui2020CoO} applied in the $xz$-plane (corresponding to an $E$ field $\sim 10^6~\mathrm{V/m}$, using the reported resistivity of $100\,\mu\Omega\,$cm~\cite{Kouvel1963Magnetic}), we find that the nonlinearly generated
staggered spin density can reach a large value $\sim 0.6\times 10^{-5}\,\mu_{\mathrm{B}}/\mathrm{nm}^{3}$. The generated N\'eel torque is
\begin{equation}
  \boldsymbol{T}^{i}=\boldsymbol{M}^{i}\times\boldsymbol{B}_{T}^{i},
\end{equation}
where $\boldsymbol{M}^{i}$ is the magnetization on sublattice $i$, and
\begin{equation}
  \boldsymbol{B}_{T}^{i}\approx-\left(J_{\mathrm{sd}}/\mu_{\mathrm{B}}\right)\delta\boldsymbol{s}^{i}/M^{i}
\end{equation}
is the effective spin-orbit magnetic field from $\delta\boldsymbol{s}^{i}$, where $J_{\mathrm{sd}}$ is the exchange coupling strength between carrier spin and background magnetization. It is noted that the $\mathcal{T}$-odd nonlinear torque here is \emph{antidamping-like}. Such a torque is more efficient than field-like torque in driving magnetic reorietation, as it competes with anisotropic barrier multiplied by the Gilbert damping $\alpha_G$, whose strength is relatively small \cite{Manchon2019Current}. To capture this character, one usually looks at the effective strength $B_T/\alpha_G$. Taking typical values $J_{\mathrm{sd}}\sim 1\,$eV and $\alpha_G\sim 0.01$~\cite{Zelezny2014Relativistic,Tang2024Lossless}, we find
$B_T^{i}/\alpha_{\mathrm{G}}\sim 180$ mT (corresponding to an energy scale $\sim 30\,\mu$eV), which is a significant value
compared to in-plane MAE and should be able to drive magnetic reorientation in MnRh.

\begin{figure}
\begin{centering}
\includegraphics[width=8.6cm]{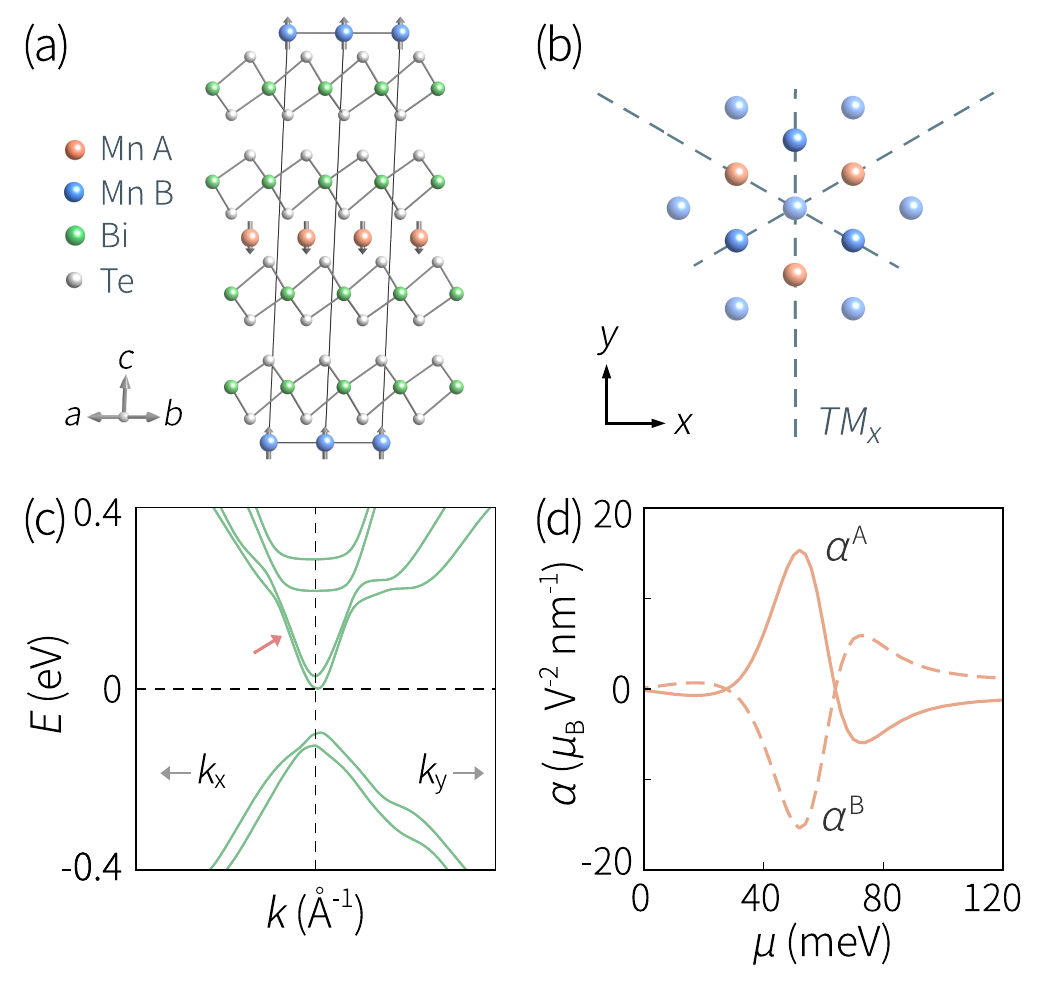}
\par\end{centering}
\caption{\protect\label{Fig_3}(a) Structure of AFM bulk MnBi$_{2}$Te$_{4}$. (b) Top view of the two Mn sublattices. Other atoms are not shown. (c) Calculated band structure of bulk MnBi$_{2}$Te$_{4}$. (d) Calculated $\alpha_{yyy}^{A}$ and $\alpha_{yyy}^{B}$ versus $\mu$ on the electron doping side. $T=20$~K is taken in the calculation of (d).}
\end{figure}

\textcolor{blue}{\textit{Application to doped 3D MnBi$_{2}$Te$_{4}$.}}
Next, we consider the 3D bulk  MnBi$_{2}$Te$_{4}$. It is an van der Waals layered AFM material, which have been attracting great interest in recent years~\cite{Otrokov2019Unique,Zhang2019Topological,Gong2019Experimental,Li2019Intrinsic,Otrokov2019Prediction,Deng2020Quantum,Liu2020Robust,Yang2021Odd,Li2023Progress,Gao2023Quantum,Wang2023Quantum,Li2024Quantum,Lian2025Antiferromagnetic,Qiu2025Observation}.
As illustrated in Fig.~\ref{Fig_3}(a), MnBi$_{2}$Te$_{4}$ consists of Te-Bi-Te-Mn-Te-Bi-Te septuple layers stacked along the $c$ axis ($z$ direction). The AFM ordering is of $A$ type, i.e., in each septuple layer, the Mn spins are ferromagnetically coupled, with an easy axis along $z$, whereas neighboring layers are coupled in AFM manner.

The AFM state has magnetic point group $-3m1^{\prime}$. It preserves $\mathcal{P}$, so linear N\'eel torque is not allowed.
In addition, $\mathcal{PT}$ and $\mathcal{T}t_{1/2}$ also exist, so $\mathcal{T}$-odd nonlinear response makes the dominant N\'eel torque. Constrained by $\mathcal{G}_s=-3m^{\prime}$, for driving field in the $xy$-plane, we find that
there is only one independent response tensor component, with $\alpha_{xxy}^{i}=\alpha_{yxx}^{i}=-\alpha_{yyy}^{i}$.

The calculated band structure of MnBi$_{2}$Te$_{4}$ is plotted in Fig.~\ref{Fig_3}(c), which shows an AFM semiconductor state. The local moment on Mn site is found to be $\sim 4.6\mu_B$. These results are consistent with previous studies~\cite{Otrokov2019Prediction}. Using Eq.~(\ref{alpha}), the nonlinear response $\alpha_{yyy}^{i}$ is evaluated and plotted in Fig.~\ref{Fig_3}(d) on the $n$-doping side.
One observes that there is a peak $\sim15\,\mu_{\mathrm{B}}\,\mathrm{V^{-2}\,nm^{-1}}$ at about 52~meV above the conduction band minimum. This can be attributed to a small-gap region in the band structure, as indicated in Fig.~\ref{Fig_3}(c), which hosts enhanced BCPs.
At this doping level, under a moderate current density of $3\times 10^{7}\,\mathrm{A/cm^{2}}$~\cite{Olejnik2018Terahertz,Klaui2020CoO},
the generated staggered spin density can reach $1.4\times10^{-4}\,\mu_{\mathrm{B}}/\mathrm{nm^{3}}$. The resulting effective field $B_{T}^{i}/\alpha_{G}$ can be as large as  $\sim 11\,$T, which is more than one order of magnitude larger than the MAE ($\sim$0.2~meV per Mn~\cite{Otrokov2019Prediction,Otrokov2019Unique}) of MnBi$_{2}$Te$_{4}$, suggesting the possibility of magnetic switching by nonlinear N\'eel torque.



\begin{figure}
\begin{centering}
\includegraphics[width=8.6cm]{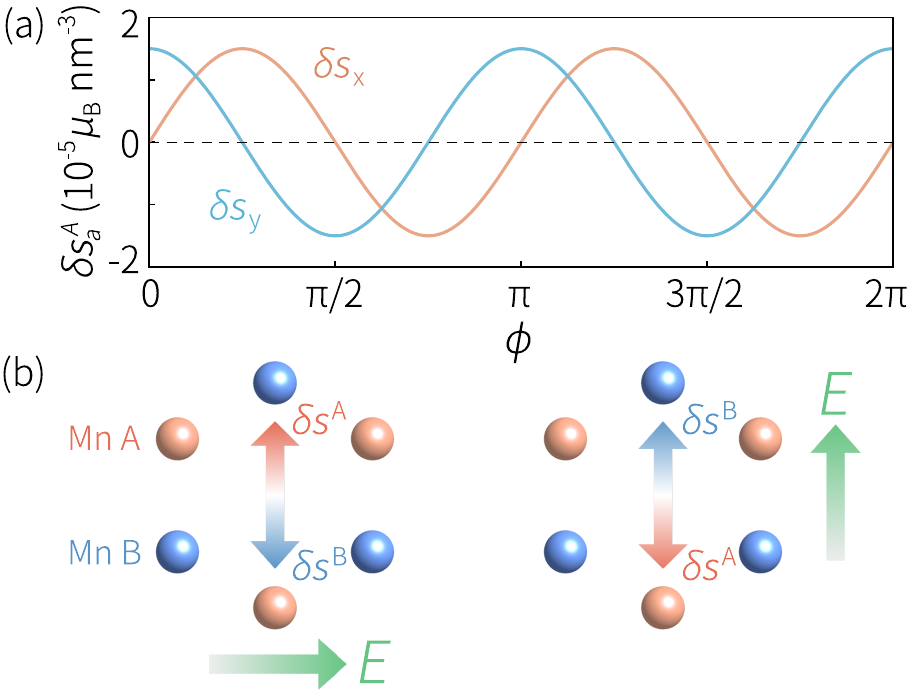}
\par\end{centering}
\caption{\protect\label{Fig_4}(a) Angular dependence of the nonlinearly generated staggered spin density in bulk MnBi$_{2}$Te$_{4}$. (b) Schematic figure showing the flip of nonlinear N\'eel torque when the driving field is rotated by $\pi/2$.}
\end{figure}

The nonlinear character of N\'eel torque is manifested in a unique angular dependence.
By rotating the driving current in the $xy$-plane, the generated spin density exhibits the following angular dependence:
\begin{equation}
  \delta\boldsymbol{s}^{i}=\alpha_{yyy}^{i}E^{2}\left({\hat{x}}\sin2\phi+{\hat{y}}\cos2\phi\right),
\end{equation}
where $\phi$ is the angle between driving current and $x$ axis.
As shown in Fig.~\ref{Fig_4}(a), the $x$ and $y$ components of the spin density have a simple $\sin 2\phi$ and $\cos 2\phi$ dependence, respectively.
One sees that the sign of the nonlinear N\'eel torque is flipped when the driving direction is changed by $\pi/2$ (Fig.~\ref{Fig_4}(b)), and it remains the same under the flip of driving current. This feature can be used to distinguish and extract the nonlinear response in experiment.

\textcolor{blue}{\textit{Application to even-layer insulating MnBi$_{2}$Te$_{4}$.}} The AFM ordering in MnBi$_{2}$Te$_{4}$ persists down to few layers~\cite{Yang2021Odd,Ovchinnikov2021Intertwined,Lujan2022Magnons}. Consider even-layer MnBi$_{2}$Te$_{4}$ in the AFM insulator state (i.e., without doping). Such systems break $\mathcal{P}$, however, $\mathcal{PT}$ symmetry is preserved. According Table~\ref{tab1}, the intrinsic nonlinear N\'eel torque is the leading effect in this case.

Symmetry analysis shows that $\alpha_{yyy}^{i}$ is still the only independent response coefficient.
Our calculation shows that for bilayer insulating MnBi$_{2}$Te$_{4}$, $\alpha_{yyy}^{i}$ can reach $\sim 0.4\,\mu_{\mathrm{B}}\,\mathrm{V^{-2}\,nm^{-1}}$. In an insulator, $E$ field can be applied through non-contact way and its magnitude can reach up to $10^8$ V/m~\citep{Olejnik2018Terahertz}. Under a moderate $E$ field of $10^{7}$ V/m,
we estimate $B_T^{i}/\alpha_G\sim 3.2\,$T ($\sim 0.8$~meV), which is more than one order of magnitude larger than the reported MAE ($\sim 0.06$~meV~\cite{Lujan2022Magnons}) in bilayer MnBi$_{2}$Te$_{4}$. The result in four septuple-layer MnBi$_{2}$Te$_{4}$ is similar, with $B_T^{i}/\alpha_G\sim 1.7\,$T.




\textcolor{blue}{\textit{Discussion.}} We have proposed the nonlinear N\'eel torque and revealed it as a new route to achieve electric control of centrosymmetric AFMs, where the existing paradigm of linear N\'eel torque fails.
The strong N\'eel torques and the induced magnetic reorientation predicted in MnRh and MnBi$_{2}$Te$_{4}$ can be readily probed in experiment, e.g., by anisotropic magnetoresistance or Hall measurement, magneto-optical detection, and magnetic resonance measurement~\cite{Klaui2020CoO,Demsar2021optical,Kent2022quantifying,Liu2022PRL}.

Our findings greatly broaden the scope of N\'eel torque and AFM spintronics.
Moreover, as noted, not just for $\mathcal{P}$-symmetric AFMs, the nonlinear mechanism may also dominate in certain $\mathcal{P}$-broken AFMs. It should also be mentioned that in systems where linear and nonlinear responses coexist, the nonlinear effect is not necessarily weaker, 
and it can in general be separated from the linear effect by symmetry, e.g., the linear torque is odd under the reversal of the driving field or current, whereas the nonlinear torque is even.

%
%
%

The formula and the calculation presented here are focused on the intrinsic contribution. In metallic systems, there also exist scattering induced
extrinsic contributions. For example, we have evaluated
the AFM nonlinear Edelstein effect, which results from the $E$-field induced second-order shift of Fermi surface. The corresponding response tensor is given by
\begin{equation}
    \alpha_{abc}^{\mathrm{Edel},i}=-\frac{\tau^{2}}{2}\int\left[d\boldsymbol{k}\right]f_{0}^{\prime}\left(v_{b}\partial_{k_{c}}+v_{c}\partial_{k_{b}}\right)s_{a}^{i},
\end{equation}
where $\tau$ is the scattering time. Our calculation finds that for MnRh and doped MnBi$_{2}$Te$_{4}$, this extrinsic contribution is orders of magnitude smaller than the intrinsic one. Nevertheless, understanding extrinsic torques in AFMs is also an important task for future studies.

%

\end{document}